\documentclass{aastex631}
\usepackage[english]{babel} 
\usepackage{amssymb}
\usepackage{amsmath}
\usepackage{txfonts}
\usepackage{mathdots}
\usepackage{graphicx}
\usepackage{subfigure}
\usepackage{multirow}
\usepackage[bottom]{footmisc}
\shorttitle{IXPE Telescope Ground Calibration}
\shortauthors{Ramsey B., Kolodziejczak J...,Fabiani, S., Muleri, F., et al.}

\begin{document}

\title{The Telescope Calibration of the Imaging X-ray Polarimetry Explorer}

\author[0000-0003-1548-1524]{Brian Ramsey}
\affiliation{NASA Marshall Space Flight Center, 
Huntsville, Al 35812, USA}
\author [0000-0002-0110-6136]{Jeffery Kolodziejczak}
\affiliation{NASA Marshall Space Flight Center,
Huntsville, Al 35812, USA}
\author [0000-0002-5106-0463]{Wayne Baumgartner}
\affiliation{NASA Marshall Space Flight Center,
Huntsville, Al 35812, USA}
\author [0000-0003-0411-4606]{Nicholas Thomas}
\affiliation{NASA Marshall Space Flight Center,
Huntsville, Al 35812, USA}
\author [0000-0002-0901-2097]{Stephen Bongiorno}
\affiliation{NASA Marshall Space Flight Center,
Huntsville, Al 35812, USA}
\author [0000-0002-3638-0637]{Phillip Kaaret}
\affiliation{NASA Marshall Space Flight Center,
Huntsville, Al 35812, USA}
\author [0000-0002-1868-8056]{Stephen O'Dell}
\affiliation{NASA Marshall Space Flight Center,
Huntsville, Al 35812, USA}
\author [0000-0002-9443-6774]{Allyn Tennant}
\affiliation{NASA Marshall Space Flight Center,
Huntsville, Al 35812, USA}
\author [0000-0002-5270-4240]{Martin C. Weisskopf}
\affiliation{NASA Marshall Space Flight Center,
Huntsville, Al 35812, USA}
\author[0000-0003-1533-0283] {Sergio Fabiani}
\affiliation{INAF-Instituto di Astrofisica e Planetologia Spaziali,
Via del Fosso del Cavaliere 100,
00133 Rome, Italy}
\author[0000-0003-3331-3794] {Fabio Muleri}
\affiliation{INAF-Instituto di Astrofisica e Planetologia Spaziali,
Via del Fosso del Cavaliere 100,
00133 Rome, Italy}
\author[0000-0002-7781-4104] {Paolo Soffitta}
\affiliation{INAF-Instituto di Astrofisica e Planetologia Spaziali,
Via del Fosso del Cavaliere 100,
00133 Rome, Italy}
\author[0000-0003-4925-8523] {Enrico Costa}
\affiliation{INAF-Instituto di Astrofisica e Planetologia Spaziali,
Via del Fosso del Cavaliere 100,
00133 Rome, Italy}
\author[0000-0003-0331-3259] {Alessandro Di Marco}
\affiliation{INAF-Instituto di Astrofisica e Planetologia Spaziali,
Via del Fosso del Cavaliere 100,
00133 Rome, Italy}
\author[0000-0003-1074-8605] {Riccardo Ferrazzoli}
\affiliation{INAF-Instituto di Astrofisica e Planetologia Spaziali,
Via del Fosso del Cavaliere 100,
00133 Rome, Italy}
\author[0000-0001-8916-4156] {Fabio La Monaca}
\affiliation{INAF-Instituto di Astrofisica e Planetologia Spaziali,
Via del Fosso del Cavaliere 100,
00133 Rome, Italy}
\author[0000-0002-9774-0560] {John Rankin}
\affiliation{INAF-Osservatorio Astronomico Di Brera,
Via E.Bianchi 26,
23807 Merate, Italy}
\author[0000-0003-0411-4243] {Ajay Ratheesh}
\affiliation{INAF-Instituto di Astrofisica e Planetologia Spaziali,
Via del Fosso del Cavaliere 100,
00133 Rome, Italy}
\author[0000-0002-3180-6002] {Alessio Trois}
\affiliation{INAF-Osservatorio Astronomico Di Cagliari,
Via dela Scienza 5,
09047 Selargius, Italy}
\author[0000-0002-9785-7726] {Luca Baldini}
\affiliation{INFN - Sezione di Pisa,
Largo B. Pontecorvo 3,
56127 Pisa, Italy}
\author[0000-0002-2469-7063] {Ronaldo Bellazzini}
\affiliation{INFN - Sezione di Pisa,
Largo B. Pontecorvo 3,
56127 Pisa, Italy}
\author[0000-0002-9460-1821] {Alessandro Brez}
\affiliation{INFN - Sezione di Pisa,
Largo B. Pontecorvo 3, 
56127 Pisa, Italy}
\author[0000-0002-0984-1856] {Luca Latronico}
\affiliation{INFN - Sezione di Torino,
Via Pietro Giuria 1,
10125 Torino, Italy}
\author[0000-0002-5916-8014] {Leonardo Lucchesi}
\affiliation{INFN - Sezione di Pisa,
Largo B. Pontecorvo 3,
56127 Pisa, Italy}
\author[0000-0002-0998-4953] {Alberto Manfreda}
\affiliation{INFN - Sezione di Napoli,
Strada Comunale Cinthia,
80126 Napoli, Italy}
\author[0000-0001-9577-2588] {Massimo Minuti}
\affiliation{INFN - Sezione di Pisa,
Largo B. Pontecorvo 3,
56127 Pisa, Italy}
\author[0000-0002-3446-8969] {Leonardo Orsini}
\affiliation{INFN - Sezione di Pisa,
Largo B. Pontecorvo 3,
56127 Pisa, Italy}
\author[0000-0003-0455-2358] {Michele Pinchera}
\affiliation{INFN - Sezione di Pisa,
Largo B. Pontecorvo 3,
56127 Pisa, Italy}
\author[0000-0001-5676-6214] {Carmelo Sgr\`{o}}
\affiliation{INFN - Sezione di Pisa,
Largo B. Pontecorvo 3,
56127 Pisa, Italy}
\author[0000-0003-0802-3453] {Gloria Spandre}
\affiliation{INFN - Sezione di Pisa,
Largo B. Pontecorvo 3,
56127 Pisa, Italy}
\begin{abstract}
Fifty years after the very first sounding rocket measurement of cosmic X-ray polarization, the Imaging X-ray Polarimetry Explorer (IXPE) mission has effectively opened a new window into the X-ray sky. Prior to launch of IXPE, an extensive calibration campaign was carried out to fully characterize the response of this new type of instrument. Specifically, the polarization-sensitive detectors were intensively calibrated in Italy, where they were developed and built. The X-ray optics, which collect and focus X rays onto the detectors, were built and calibrated in the U.S. A key question was whether the telescope (optics + detectors) calibrations could be synthesized from the individual component calibrations, avoiding time consuming and costly end-to-end calibrations for a flight program with a fixed schedule. 

The data presented here are from a calibration of the flight spare telescope utilizing the flight spare detector and flight spare mirror assembly combined. These data show that the presence of the mirror module does not affect the polarization response of the detectors (within the required calibration accuracy) and that the angular resolution of the telescopes could be accurately determined. Thus, the original extensive stand-alone ground calibration data of all the flight detectors and all the flight optic can be utilized in full to derive the flight telescopes calibrations.

\end{abstract}

\section{Introduction}
\label{sec:intro} 
\begin{figure}[ht]
\label{fig:IXPE} \centering
\includegraphics[width=6.5in, height=3.5in, keepaspectratio=true]{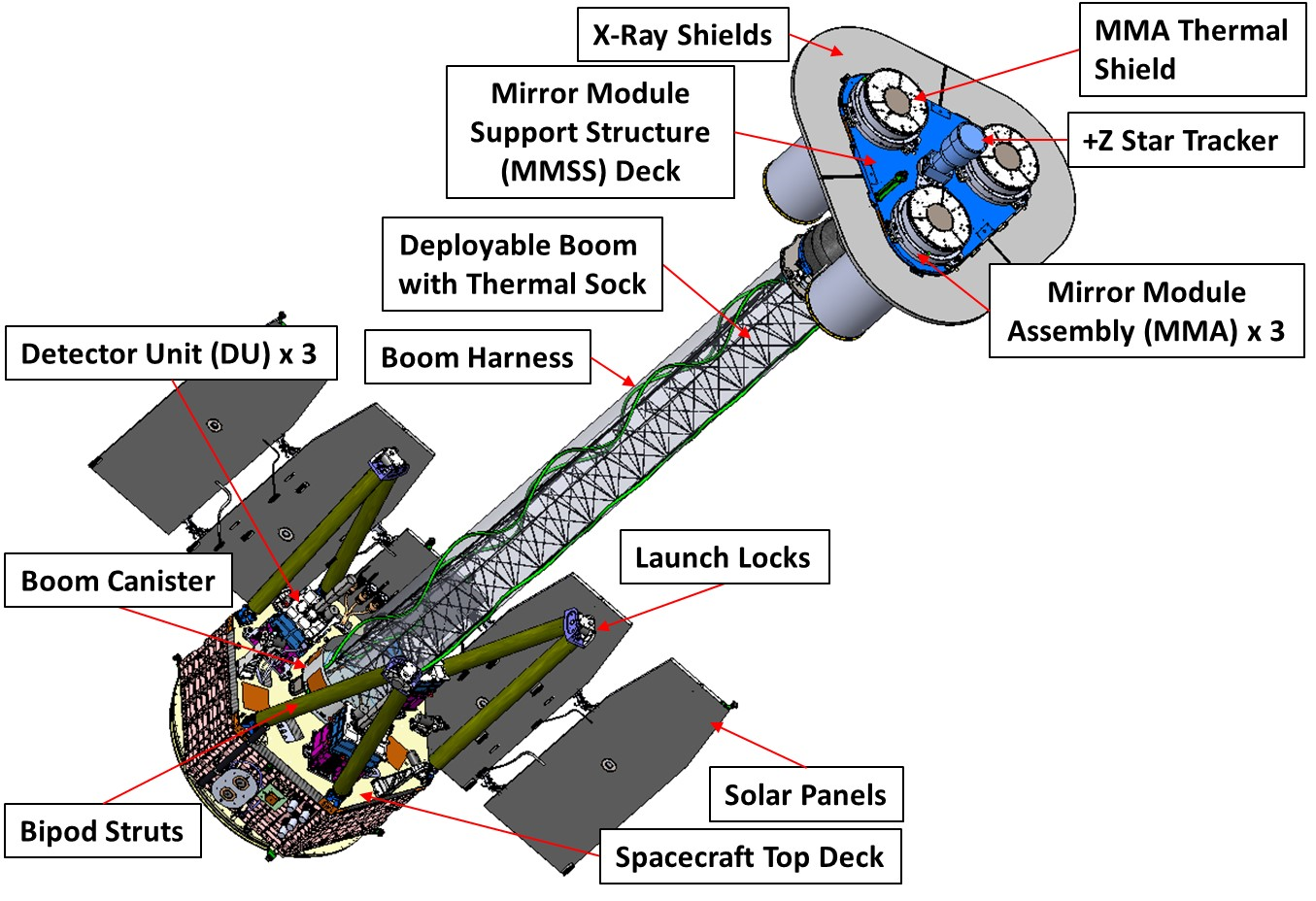}
\caption{\textit{The Imaging X-ray Polarimetry Explorer with three
polarization-sensitive Detector Units at the focus of three X-ray Mirror Module Assemblies}.}
\end{figure}

Launched in late 2021, IXPE, the Imaging X-ray
Polarimetry Explorer (IXPE, Figure \ref{fig:IXPE}) is a NASA
small explorer mission that has opened a new window on the X-ray
sky, 50 years since the first observations were made \citep{Novick1972, Novick1978}. The IXPE observatory, placed in a low-earth (600 km)
equatorial orbit, consist of 3 identical telescopes each
comprising an X-ray mirror module assembly (MMA) with a
polarization-sensitive imaging X-ray detector unit (DU) at its
focus. An extending boom, deployed on orbit, provides the
necessary 4-m focal length and the payload sits atop a 3-axis
stabilized spacecraft that provides for the payload command and
control. IXPE is a collaboration between NASA and
ASI, the Italian Space Agency. In the U.S., NASA's Marshall Space
Flight Center (MSFC) is the lead Institution providing overall
program management, system engineering, safety and mission assurance oversight, and science operations. It is also responsible for development of the
mirror module assemblies and for providing MMA and Telescope
calibrations. ASI is responsible for providing the
instrument \citep{Soffitta2021}, composed of the
polarization-sensitive imaging detectors \citep{Baldini2021} and a detectors service unit, and the
IXPE ground station situated near the equator in Malindi, Kenya. The
IXPE mission in Italy is led by the Istituto di Astrofisica e Planetologia Spaziali - Istituto Nazionale di Astrofisica (INAF-IAPS), which designed and
built the flight calibration sources \citep{Ferrazzoli2020} and
was responsible for the ground
calibration of the flight detectors \citep{Muleri2022,Dimarco2022c}. The detectors were were designed and developed at the Istituto Nazionale di Fisica Nucleare (INFN) in Pisa. Detailed
descriptions of the IXPE program can be found in
\cite{Weisskopf2022} and \cite{Weisskopf2022a}.The latter includes a discussion of IXPE science.

\section{Telescope Details}
\label{sec:Telescope details}
\subsection{Mirror Module Assembly}
\label{sec:MMA} 
\begin{figure}[ht]
 \centering
\includegraphics[width=5.1in, height=2.83in, keepaspectratio=true]{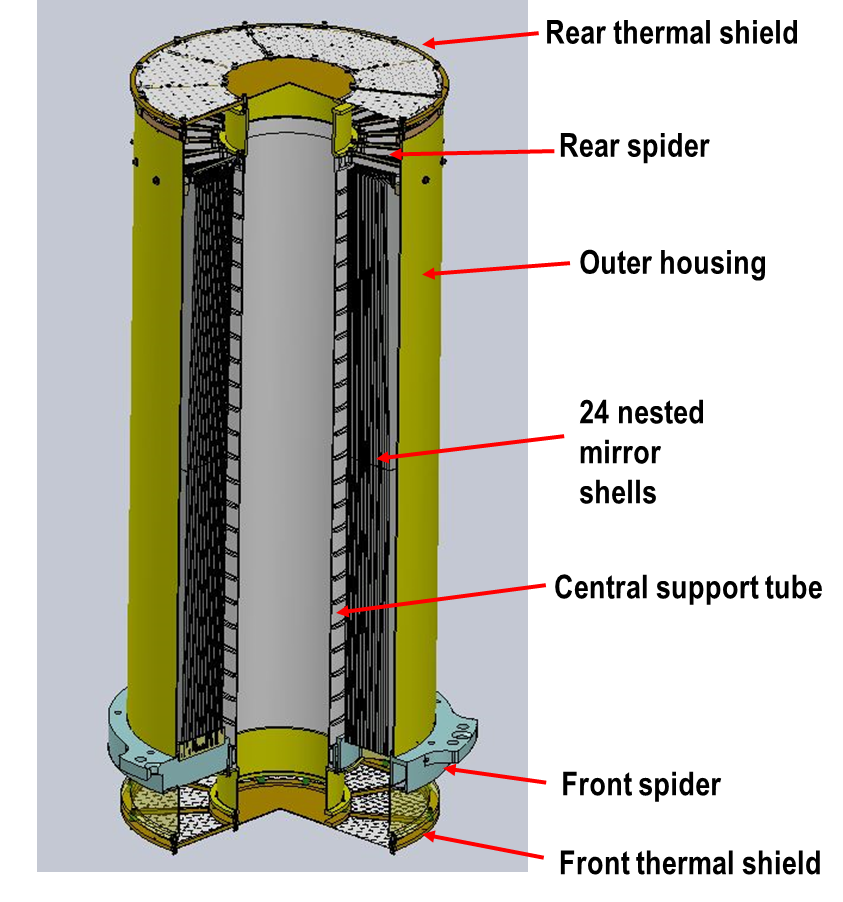}
\caption{\textit{An IXPE Mirror Module Asssembly}.}
\label{fig:mirror}
\end{figure}

Each Mirror Module Assembly (MMA,
Figure \ref{fig:mirror}) consists of 24
concentrically-nested X-ray-mirror shells fabricated via an
electroformed-nickel-replication process. Each shell contains both
the primary and secondary (parabolic and hyperbolic) components of
a Wolter-1-type optical configuration. The mirror shells are closely
packed and quite thin, to meet a science-imposed effective area
requirement while keeping the mass within launch requirements. The
shells are held in place by a forward spider into which they are
glued during shell alignment and assembly. A thin outer housing 
provides mechanical protection and thermal control via surface-mounted heaters. Completing the design
is a pair of ultra-thin thermal shields, mounted each end, that together with the housing heaters permit active control
of the MMA operating temperature while allowing X-rays in IXPE's
operating band (2-8 keV) to enter and leave with high efficiency. Table \ref{Tab:MirrorChar} 
lists the MMA parameters. Full details
of the IXPE MMA are given in \cite{Ramsey2022}.

\begin{table}[ht!]
\centering
\caption{\textit{IXPE MMA Parameters}.}
\begin{tabular}{|l|l|}
\hline
\textbf{Parameter} & \textbf{Value}\\
\hline
\text{Number of mirror modules} & \text{3}\\
\hline
\text{Number of shells per mirror module} & \text{24}\\
\hline
\text{Focal length} & \text{4 m}\\
\hline
\text{Total shell length} & \text{600 mm}\\
\hline
\text{Range of shell diameters} & \text{162–272 mm}\\
\hline
\text{Range of shell thicknesses} & \text{0.18–0.25 mm}\\
\hline
\text{Shell material} & \text{Electroformed nickel–cobalt alloy}\\
\hline
\text{Effective area per mirror module} & \text{166 cm$^{2}$ (@ 2.3 keV)}\\ 
                                           & \text{$>$ 175 cm$^{2}$ (3–6 keV)}\\
\hline
\text{Angular resolution (HPD)} & \text{$\leq$ 28 arcsec} \\
\hline
\text{Field of view (detector limited)} & \text{12.9 arcmin square}\\
\hline
\end{tabular}
\label{Tab:MirrorChar}
\end{table}

\subsection{Detector Unit}
\label{sec:DU} 
\begin{figure}[ht]
\centering
\includegraphics[width=5.1in, height=2.83in, keepaspectratio=true]{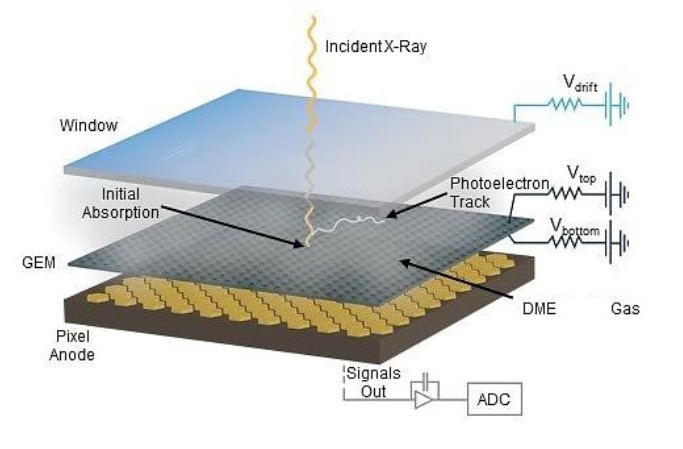}
\caption{\textit{Schematic of the Gas Pixel Detector.}}
\label{fig:GPD}
\end{figure}

The IXPE Detector Unit (DU) measures the energy,
position, arrival time and polarization of each absorbed X-ray
photon that was focused by the MMA. The heart of each DU is the
Gas Pixel Detector (GPD, Figure \ref{fig:GPD}), a small (15 mm x 15 mm)
proportional counter with a special fill gas, Dimethyl-Ether (DME), and
a fine pixel readout, via a custom Application Specific Integrated Circuit (ASIC), 
that allows the imaging
of photoelectron-induced tracks produced when X rays photoelectrically interact with the
fill medium (see \cite{kaaret2021} for a general description of the workings of photoelectric polarimeters). The initial direction of the photoelectron, the
interaction point and the total charge in the track, provide the
necessary information to determine the polarization, location and
energy of the absorbed X-ray photon, respectively. X rays enter 
through a beryllium window (Figure \ref{fig:GPD})
and interact with the DME in an absorption and drift region. The
resultant photoelectron track is then drifted down though a Gas
Electron Multiplier (GEM) where each charge in the track is
amplified before being registered on a highly-pixelated anode
plane. This anode plane is the front end of the custom ASIC which
processes the event so that it can be sent to the ground for track
reconstruction and subsequent analysis. This analysis, among other
things, bins the data to determine the polarization degree and
polarization angle for each cosmic X-ray source observed. Parameters of the
IXPE GPDs are given in Table \ref{tab:GPD}; comprehensive descriptions of
the Instrument and its GPD, can be found in \citep{Soffitta2021} and
\citep{Baldini2021}, respectively. 

\begin{table}
\caption{\textit{Gas Pixel Detector parameters.}}
\centering
\begin{tabular}{| l l |}
\hline
\textbf{Parameter} & \textbf{Value} \\
\hline
\text{Sensitive area} & \text{15 × 15 mm (13 x 13 arcmin) } \\
\hline
\text{Fill gas and fill pressure} & \text{DME @ 0.8 atmosphere} \\
\hline
\text{Detector window} & \text{50-$\mu$m thick beryllium} \\
\hline
\text{Absorption and drift region depth} & \text{10 mm} \\
\hline
\text{Spatial resolution (FWHM)} & \text{$\leq$ 123 $\mu$m (6.4 arcsec, 2 keV)} \\
\hline
\text{Energy resolution (FWHM)} & \text{0.57 keV @ 2 keV ( $\propto$ $\sqrt{E}$)} \\
\hline
\text{Useful energy range } & \text{2 - 8 keV} \\
\hline
\end{tabular}
\label{tab:GPD}
\end{table}

\section{ON-GROUND TELESCOPE CALIBRATION} \label{sec:TelescopeCalib}

\subsection{Requirements} \label{sec:Requirem}
The sensitivity of a polarimeter is
given by the equation:

\begin{equation}\label{eq:MDP}
    MDP = \frac{429 \%}{\mu} \times \frac{\sqrt{(R_{s} + R_{b})\times t}}{R_{s}\times t}
\end{equation}

\noindent where R$_{s}$ is the source count rate (a product of the source
flux, the mirror module effective area, and the detector quantum
efficiency), R$_{b}$ is the background rate, t is the observation time and $\mu$ is the modulation
factor which describes the response of the polarimeter to 100\%
polarized X rays. The sensitivity, MDP, is the minimum detectable
degree of polarization that can be measured at the 99$\%$
confidence level. The parameters $\mu$, R$_{s}$ and R$_{b}$ are all energy dependent. To accurately determine the sensitivity of the
polarimeter, the effective area of the optics must be measured
both on axis and for a variety of off-axis angles and azimuths, for
a range of energies covering the full IXPE operating band. For the
detectors, the quantum efficiency and the modulation factor must
be accurately calibrated at a series of energies covering the IXPE
band. Further, as IXPE
is designed to be sensitive to small degrees (few $\%$) of polarization it is vital to
test with unpolarized X rays to verify that there are no
systematic effects that could give rise to spurious modulations
and hence false polarizations. In fact, during detector
calibration, by far the largest amount of time was spent analyzing
the DU's response to unpolarized X rays. Finally, the spatial resolution of the detectors and the angular resolution of the optics must be measured at a range of energies and off-axis positions.

The requirements for IXPE
calibration are based on a flow-down from the top-level science
requirement to measure polarization down to a certain level for a
representative sample of various classes of cosmic X-ray sources.
This top-level polarization-sensitivity requirement is stated as:

\emph{IXPE shall provide a minimum detectable polarization
(MDP$_{99}$) not to exceed 5.5$\%$ for a point source with an
E$^{-2}$ photon spectrum and a 2--8 keV flux of 10$^{-11}$ ergs
cm$^{-2}$ sec$^{-1}$ and an integration time of 10 days.}

This describes the level of polarization that can be measured at
the 99$\%$ confidence level for a generic cosmic source spectrum,
flux and integration time. A second top-level requirement concerns
the angular resolution of the observatory, to perform its imaging
role:

\emph{IXPE shall have a system-level angular resolution not to
exceed 30 arcsec half-power diameter (HPD)}.

These two requirements in turn flow down to individual
requirements on telescope effective area, telescope modulation
factor, telescope spurious modulation and telescope angular
resolution, all at multiple energies across the IXPE band. Derived
from these, in turn, are the calibration requirements that describe the
precision to which these values must be measured. These telescope
on-ground calibration requirements are as follows:

\begin{enumerate}
    \item The effective area of each MMA-DU combination shall be measured to an accuracy equal to or better than 10\% of its value, on axis and at least 16 off-axis positions, for at least 3 energies between 1.5 and 8 keV.
    \item The modulation factor ($\mu$) of each MMA-DU combination shall be measured to an accuracy equal to or better than 1\% of its value for at least 2 energies in the range 1.5 – 8.0 keV and with at least 2 dither patterns at each energy.
    \item The spurious-modulation amplitude of each MMA-DU combination shall be measured to an accuracy equal to or better than 0.1\% absolute value, for at least 2 energies in the range 1.5 – 8 keV and with at least 1 dither pattern at each energy.
    \item The half-power diameter (HPD) of each MMA-DU combination shall be measured to an accuracy equal to or better than 3\% of its value, on axis and at least 16 off-axis positions, for at least 3 energies between 1.5 and 8 keV.
\end{enumerate}

The mention of dither pattern in the above requirements refers to
a controlled small-amplitude (few arcmin) variation of observatory pointing planned to distribute the
focused image over a small (few mm) region of the focal plane - this
removes the requirement for accurately calibrating the
response of every pixel in the detector. The dither pattern is designed to uniformly illuminate the detector over the dithered region. 

Extensive calibrations were
carried out on the MMAs at MSFC and the DUs at INAF-IAPS, Italy. The response of a telescope system could then, in theory, be synthesized by simply
combining the responses of the individual optics and detectors.
However, there were concerns that there were small differences in
the functioning of a telescope system compared with how the
individual components were tested. The most significant of these
was that the detector units were all calibrated with near-parallel beams of X rays oriented normally to the detector
surface. However, when placed at the focus of an X-ray optic, the X rays
are now impinging in a cone, focused to the center of the detector
depth, at angles that are 4 $\times$ the on-axis graze angles of the
optics. For the outer shells, which provide the bulk of the
effective area, these impingement angles are about 2 degrees. This
affects both the spatial resolution, which is now somewhat blurred
due to parallax effects, and potentially the modulation factor due
to a tipping of the photoelectron emission plane. 
Also, the reflection of X rays can generate polarization and
though this effect is calculated to be small for typical graze
angles ($<0.01\%$ polarization for a 0.5-degree graze angle following the formalism in \citet{Almeida1992}), and reduced by azimuthal symmetry, it is important to demonstrate that it does not contribute
at any IXPE energy or possible off-axis angle. Finally, during DU
testing it was found that spurious modulation did occur with
unpolarized X rays, the cause of which was never fully understood 
although it was found to originate in the GEM.
While this spurious modulation was stable and could therefore be
calibrated out (it varied with energy and position on the
detector), it was important to show that it remained the same
when the detector was at the focus of an MMA, and hence that the
extensive individual detector calibrations with unpolarized X rays
were still valid. 

The original intention was to perform a full
end-to-end calibration of each telescope system, hence the above individual 
telescope calibration requirements. However, as the
program progressed, and various hurdles were overcome (including Covid-19 and a US
Government shutdown), it became obvious that there was not enough
time and funding to accomplish this, given a fixed launch date. It was therefore decided that the flight
spare telescope would be fully calibrated to prove that we could
accurately perform a telescope calibration by the appropriate analytical
combination of the individual flight spare mirror module assembly
(MMA-4) and the flight-spare detector (DU-FM1) calibrations. This would pave the way for analytical combinations for all flight telescopes by demonstrating that any effects induced by the presence of the optics were
negligible within the required calibration accuracy. In this manner, the required response functions for all three flight telescopes could be confidently generated. \footnote{\url{https://heasarc.gsfc.nasa.gov/docs/ixpe/analysis/IXPE-SOC-DOC-011A_UG-Observatory.pdf}}

\section{MSFC Calibration Facility Modifications.} \label{subsec:CalFac}

\begin{figure}[htb!] \centering
\subfigure[\label{subFig:MIRROR@SLTF.png}]{\includegraphics[height=6.0
cm]{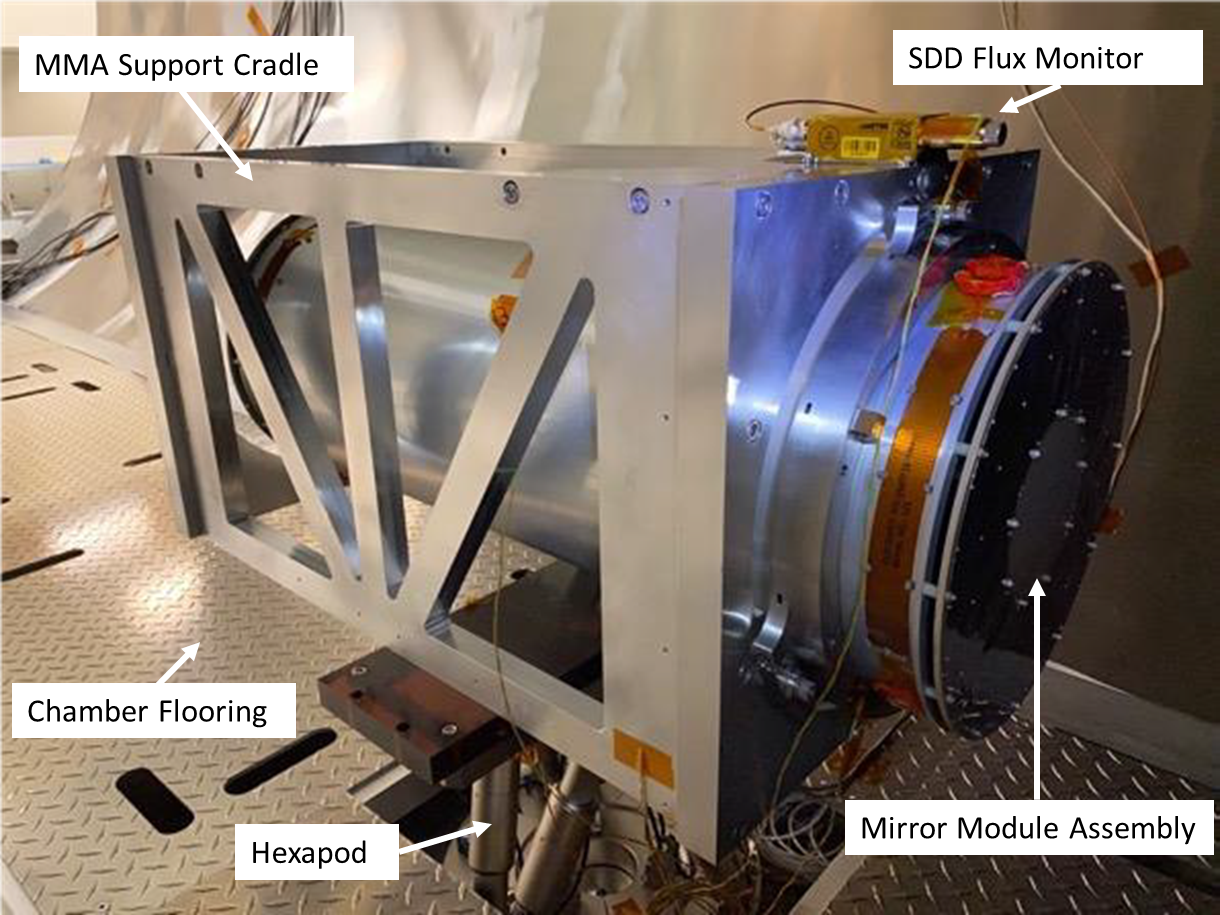}}
\subfigure[\label{subFig:DetFac&SLTF}]{\includegraphics[height =
6.0 cm]{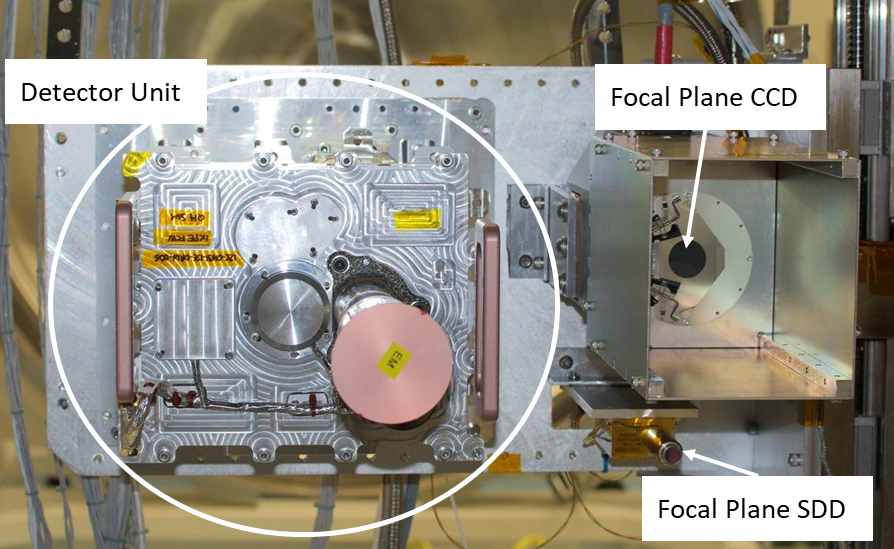}} \label{subFig:SLTF} \caption{\textit({\bf a})
\textit{IXPE MMA in test facility. ({\bf b}) Focal plane test detectors including a
CCD camera, a fast silicon drift detector (SDD) and the
flight-spare IXPE detector unit (circled)}.}
\end{figure}

All telescope (and MMA) calibrations were carried out at the MSFC
stray light test facility (now renamed the MSFC 100-m facility).
As the name implies, the facility has a $\sim$ 100-m-long vacuum
beam-line consisting of a 1.2-m-diameter beam tube feeding into
2.4-m-diameter by 14-m-long test chamber \citep{Thomas2023}. The test IXPE-spare MMA
(MMA-4) was mounted in a cradle atop a hexapod which permited fine
control in all axes (see Figure \ref{subFig:MIRROR@SLTF.png}). The IXPE spare flight detector
and facility detectors, were mounted on an XY stage system at an
appropriate distance for the 100-m source distance which gave an
effective focal length of 4.17 m. (Figure \ref{subFig:DetFac&SLTF}).

\begin{figure}
    \centering
    \includegraphics[width=0.5\linewidth]{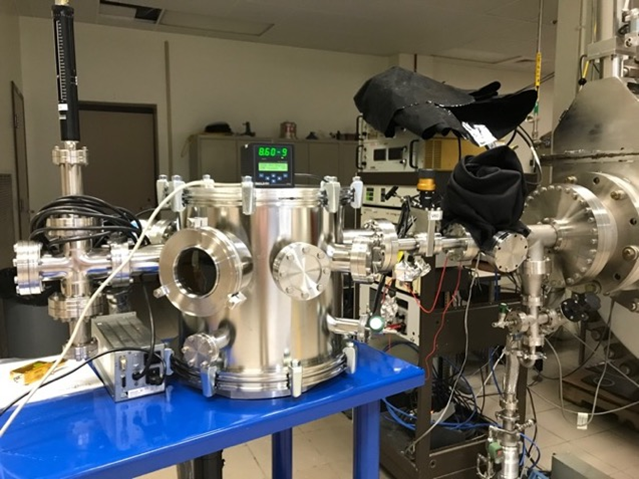}
    \caption{\textit{The test facility crystal box}.}
    \label{fig:CrystalBox}
\end{figure}

The facility contains the typical selection of electron-impact
X-ray sources with tungsten, molybdenum,  iron, titanium and
aluminum anodes giving characteristic lines and continuum
radiation. It also has an X-ray filter system to reduce continuum
emission and emphasize line emission when needed. For the IXPE telescope
tests, it was also necessary to provide highly-polarized X rays
covering the IXPE energy band.  This was done by using crystals
matched to specific X-ray lines satisfying the Bragg condition at
$\sim$ 45 degrees incidence and thus giving $>99\%$ polarization.
Table \ref{tab:Crystals} shows the configurations used. To give adequate flux it
was necessary to procure custom X-ray sources; windowless
devices, water cooled with 1-mm X-ray spot sizes and capable of operating at up to 900 Watts
power. As they need to be operated at high vacuum ($< 10^{-5}$
torr), and to block visible light emitted by the sources, an optical
blocking filter was installed. This permitted the source side to
be kept at a lower pressure than the overall facility. The whole
source assembly, with its own pumping system was mounted inside a
crystal 'box' which allowed for translation and rotation of the
different crystal assemblies (Figure \ref{fig:CrystalBox})

\begin{table}[ht!]
\centering
\caption{\textit{Polarized source configurations}.}
\begin{tabular}{| c c c c |}
\hline
\textbf{X-ray Tube (Line)} & \textbf{Crystal} & \textbf{Energy} & \textbf{Polarization} \\
\hline
\text{Rhodium (L)} & Germanium (111) & 2.7 keV & $>99\%$  at $\sim$45$^\circ$ \\
\hline
\text{Titanium (K)} & Silicon (220) & 4.5 keV & $>99\%$  at $\sim$45$^\circ$ \\
\hline
\text{Iron (K)} & Silicon (400)& 6.4 keV & $>99\%$  at $\sim$45$^\circ$ \\
\hline
\end{tabular}
\label{tab:Crystals}
\end{table}

The IXPE program purchased two identical Amptek Fast Silicon Drift Detectors (SDD)
for use during the IXPE MMA calibrations, one detector serving as a
beam monitor to measure the flux incident on the X-ray optic under
test, and the other mounted for use as a focal plane detector to
provide effective area measurements. The SDDs have a
circular aperture of area 50 mm$^{2}$, and a Be window of nominal
thickness 0.0005 inch (12.7 $\mu$m). These detectors were operated
with shaping-times of 1 microsecond or faster to accommodate
high count rates with dead-times less than 1$\%$, at the expense of
some slight degradation in spectral resolution (but still easily
meeting IXPE requirements). For characterizing the MMA Point Spread 
Function an Andor DW436 camera was used, having a 2048$\times$2048 Charge Coupled Device (CCD) sensor
with 13.5-$\mu$m pixels. The DW436 is water cooled and designed to
work entirely in vacuum. This camera is a back-illuminated, soft-X-ray-sensitive design with no coatings. The CCD camera has an optical
blocking filter (aluminized polyimide) at its entrance and uses a
Uniblitz X-ray shutter system to control integration times.

\section{Results} \label{subsec:Results}
\subsection{Effective Area}
\subsubsection{Calibration Requirement} \label{sec:EA}
The effective area of the MMA-DU combination shall be measured to
an accuracy equal to or better than 10$\%$ of its value on axis and at least 16
off-axis positions, for at least 3 energies between 1.5 and 8 keV.
\subsubsection{Measurements}
Prior to making measurements, the DU was centered on the MMA beam and a best-focus
position was obtained by minimizing the measured MMA half-power
diameter. Angular resolution (half-power diameter) and effective
area data were collected at the same time as MMA-4 was stepped
through the required on axis and off-axis angles. As rotating the
optic about its node does not move the focal image the detector
must also be stepped to appropriate off-axis positions to simulate
in-flight telescope performance. Table \ref{tab:OpticsPosition} shows the angles covered, and their
respective nominal detector positions. These angles align with a
subset of those used in MMA calibrations, allowing direct
comparisons. It is worth noting that, on orbit, off-axis pointing will change the incidence angles on the detector, but as the total field of view of the telescope is quite small (7arcmin radius) compared to the 2 degree cone angle, this effect is negligible. 

\begin{table}[ht!]
\centering
\caption{\textit{Off-axis angles and detector positions for telescope effective area measurements}.
}
\begin{tabular}{| c c c c |}
\hline
\multicolumn{2}{|c}{\textbf{(MMA)}} & \multicolumn{2}{c|}{\textbf{Detector Offset}}\\
\multicolumn{2}{|c}{\textbf{(arcmin)}} & \multicolumn{2}{c|}{\textbf{(mm)}}\\
\textbf{Pan (X)} & \textbf{Tip (Y)} & \textbf{X} & \textbf{Y}\\

\hline
0 & 0 & 0 & 0 \\
\hline
0 & 5.0 & 0 & 6.05 \\
\hline
0 & -5.0 & 0 & -6.05 \\
\hline
5.0 & 0 & 6.05 & 0 \\
\hline
-5.0 & 0 & -6.05 & 0 \\
\hline
4.95 & 4.95 & 5.99 & 5.99 \\
\hline
4.95 & -4.95 & 5.99 & -5.99 \\
\hline
-4.95 & 4.95 & -5.99 & 5.99 \\
\hline
-4.95 & -4.95 & -5.99 & -5.99 \\
\hline
0 & 3.0 & 0 & 3.63 \\
\hline
0 & -3.0 & 0 & -3.63 \\
\hline
3.0 & 0 & 3.63 & 0 \\
\hline
-3.0 & 0 & -3.63 & 0 \\
\hline
2.83 & 2.83 & 3.42 & 3.42 \\
\hline
2.83 & -2.83 & 3.42 & -3.42 \\
\hline
-2.83 & 2.83 & -3.42 & 3.42 \\
\hline
-2.83 & -2.83 & -3.42 & -3.42 \\
\hline

\end{tabular}
\label{tab:OpticsPosition}
\end{table}

Data were taken with Mo L, Ti K, and Fe K characteristic X rays.
To suppress continuum emission each source was filtered using Nb
(5 $\mu$m), Ti (77 $\mu$m) and Fe (77 $\mu$m) respectively. A
monitor counter, positioned at the MMA, was used to derive the
incident flux. This SDD has precisely-known collecting area and quantum efficiency. For each
energy and angle, a minimum of 1000 counts were collected in the
monitor counter to give the desired effective-area accuracy. As
the energy resolution of the DU is about 17$\%$ FWHM at 6.4 keV,
it cannot resolve the K$_{\alpha}$ from the K$_{\beta}$ lines for
either Ti or Fe, or the different L lines from molybdenum;
therefore, a weighted mean energy was used for these measurements.
The data from the monitor SDD, which has high spectral resolution,
is the sum all the K lines for each source. The L lines from
molybdenum are not resolved by the SDD. The effective weighted
energies for the three X-ray sources, correcting for SDD
efficiency, are: 2.30 keV for Mo L, 4.63 keV for Ti K and 6.58 keV
for Fe K.
\subsubsection{Results}\label{sec:EAMFMeas}
The data were processed using a standard moments analysis \citep{Baldini2021}.
Because the gas detector has limited energy resolution and has
tailing - lower energy events present due to incomplete charge
collection - the acceptance band of the DU was opened to + 3$\sigma$
above the peak pulse height and also included everything below the peak.
Similarly, as discussed above, the acceptance band of the SDD detector was opened to
include all K$_{\alpha}$ and K$_{\beta}$ lines plus any escape peaks (from silicon fluorescence escaping the detector) or tailing (from incomplete charge collection)
present in the data. The DU data were corrected for deadtime and the resulting count
rate was divided by the rate in the SDD flux monitor (the same one
that was used for MMA calibration), and multiplied by the
flux-monitor area. Finally, the resulting effective areas were
scaled to that for infinite source distance, using energy and
off-axis-angle dependent correction factors derived from ray-trace models.

The resulting data are shown below for each energy in Tables \ref{tab:MMEffAreaComp}, \ref{tab:EffArea2.3keV}, \ref{tab:EffArea4.6keV}, \ref{tab:EffArea6.6keV}. The uncertainties for each measurement (statistical plus
uncertainties in detector response and geometry) are given in 
table captions \ref{tab:EffArea2.3keV},\ref{tab:EffArea4.6keV},\ref{tab:EffArea6.6keV}. 

The on-axis effective area for the MMA alone was
measured during MMA calibration using a pair of identical, fast
SDDs, one (near the MMA entrance
aperture) monitoring the input flux and the other at the mirror
focus. The ratio of the two count rates, adjusted for minor
response differences measured during cross calibration, multiplied
by the monitor's active area gave the effective area of the optic
as a function of energy. 

\begin{table}[ht!]
\centering
\caption{\textit{Comparison of predicted and measured telescope on-axis effective areas. Difference column is (measured-predicted)/average value, expressed as a percentage}.
}
\begin{centering}
\begin{tabular}{| c c c c c c c |}
\hline
\textbf{Source} & \textbf{Mean Energy} & \textbf{Measured MMA} &  \textbf{DU quantum} & \textbf{Predicted} &  \textbf{Measured} &  \textbf{Difference} \\
                & \textbf{$K_{\alpha}$}+$K_{\beta}$ (keV) & \textbf{Effective Area} & \textbf{efficiency} &  \textbf{Telescope Effective} & \textbf{Effective Area} & \textbf{$\%$}\\
                &                            &   \textbf{(cm$^{2}$)} &   &  \textbf{Area (cm$^{2}$}) & \textbf{(cm$^{2}$)} & \\
\hline
\text{Mo} & 2.30 & 163.2 & 0.155 & 25.23 & 25.34 & 0.4 \\
\hline
\text{Ti} & 4.63 & 195.7 & 0.0382 & 7.48 & 7.39 & -1.3 \\
\hline
\text{Fe} & 6.58 & 176.1 & 0.0152 & 2.68 & 2.82 & 4.9
\\
\hline
\end{tabular}
\end{centering}
\label{tab:MMEffAreaComp}
\end{table}

To derive a predicted telescope on-axis
effective area from separate MMA and DU calibrations, the MMA
effective area (corrected to infinite source distance) was
combined with the DU quantum efficiency (QE) appropriate for the telescope
calibration at MSFC. This required adjusting the efficiency
measured during the DU calibration at IAPS, to correct for a
small (few mbar) pressure drop (thought to be due to absorption of the DME by epoxy used in the GPD contruction) between that date and the calibration date at MSFC,
using empirical models developed in Italy \citep{Baldini2021}. It is
also very important that the appropriate weighted mean source
energy is used when deriving the appropriate DU quantum efficiency
as the QE changes with energy. 

The results of this comparison
between the predicted on-axis telescope effective area and the
measured effective area, are shown in Table \ref{tab:MMEffAreaComp} at 3 energies with
both data sets corrected to infinite source distance. Statistical
and systematic errors are approximately 3$\%$ for the predicted
effective area (1-$\sigma$) and 5$\%$, 4$\%$ and 2$\%$ respectively (1-$\sigma$)
for the measured effective areas at 2.3 keV, 4.6 keV and 6.6 keV
respectively. It can be seen that the predicted and measured telescope 
effective areas are
within statistical errors (at 1.3-$\sigma$ for the 6.6 keV case) for these on-axis measurements, as
expected.
The same process was repeated at each energy for all off-axis
angles in Table \ref{tab:OpticsPosition}. For these, MMA effective areas 
were measured at a subset of angles using the SDDs (17 angles at 2.3 keV and 5
angles at 4.6 keV and 6.6 keV), then interpolated using CCD data
taken at every angle. The resulting effective area data are shown
in Tables \ref{tab:EffArea2.3keV}--\ref{tab:EffArea6.6keV}.

\begin{table}[htp]
\centering
\caption{\textit{Comparison of measured and predicted effective areas (EA) at 2.3 keV. Effective area 1-$\sigma$ uncertainties are $\sim$ $\pm$3\% for predicted values and $\pm$5\% for measured values. Difference column is (measured-predicted)/average value, expressed as a percentage. The mean difference value over the data set is -5.0\%, with a standard deviation of 4.1\%}.}
\begin{tabular}{| c c c c c |}
\hline
\multicolumn{2}{|c}{\textbf{MMA}} & \textbf{Predicted} & \textbf{Measured} & \textbf{Difference}\\
\multicolumn{2}{|c}{\textbf{(arcmin)}} & \textbf{Telescope} & \textbf{Telescope} & \textbf{$\%$}\\
\textbf{Pan (X)} & \textbf{Tip (Y)} & \multicolumn{1}{r}{\textbf{EA (cm\textsuperscript{2}) }} &  \multicolumn{1}{r}{\textbf{EA (cm\textsuperscript{2}) }} & \\
\hline
0 & \multicolumn{1}{c}{0} & 25.2 & 25.3 & 0.4 \\
\hline
0 & \multicolumn{1}{c}{5.0} & 19.3 & 18.9 & -2.5 \\
\hline
0 & \multicolumn{1}{c}{-5.0} & 19.3 & 17.7 & -8.6 \\
\hline
5.0 & \multicolumn{1}{c}{0} & 19.2 & 18.1 & -5.9 \\
\hline
-5.0 & \multicolumn{1}{c}{0} & 19.2 & 19.0 & -1.1 \\
\hline
4.95 & \multicolumn{1}{c}{4.95} & 16.7 & 17.0 & 1.3 \\
\hline
4.95 & \multicolumn{1}{c}{-4.95} & 16.6 & 14.6 & -13.2 \\
\hline
-4.95 & \multicolumn{1}{c}{4.95} & 16.6 & 15.7 & -5.7 \\
\hline
-4.95 & \multicolumn{1}{c}{-4.95} & 16.7 & 15.4 & -8.4 \\
\hline
0 & \multicolumn{1}{c}{3.0} & 22.2 & 21.4 & -3.8 \\
\hline
0 & \multicolumn{1}{c}{-3.0} & 22.2 & 20.9 & -5.9 \\
\hline
3 & \multicolumn{1}{c}{0} & 22.1 & 19.9 & -10.5 \\
\hline
-3 & \multicolumn{1}{c}{0} & 22.1 & 22.4 & 1.0 \\
\hline
2.83 & \multicolumn{1}{c}{2.83} & 20.7 & 20.0 & -3.1 \\
\hline
2.83 & \multicolumn{1}{c}{-2.83} & 20.6 & 19.1 & -7.6 \\
\hline
-2.83 & \multicolumn{1}{c}{2.83} & 20.6 & 19.2 & -7.2 \\
\hline
-2.83 & \multicolumn{1}{c}{-2.83} & 20.7 & 19.7 & -4.9\\
\hline
\end{tabular}
\label{tab:EffArea2.3keV}
\end{table}
\subsubsection{Discussion}\label{sec:EADisc}
It is clear that there is reasonable agreement between the predicted and measured off-axis effective areas for all energies tested. The captions for Tables \ref{tab:EffArea2.3keV},\ref{tab:EffArea4.6keV},\ref{tab:EffArea6.6keV} give the mean difference and standard deviation for all the off-axis measurements at each specific energy. These mean values show there is a small energy-dependent offset that is around -5$\%$ at 2.3 keV, -3 $\%$ at 4.6 keV and 1.5$\%$ at 6.6 keV. The standard deviation around these means is consistent with the estimated uncertainties in the measurements.These small offsets, decreasing in absolute amplitude with energy, are well within the required calibration accuracy, and are thought to be due to a possible error in the energy-dependent quantum efficiency of the detector calculated for the time of calibration.

\begin{table}[htp]
\centering
\caption{\textit{Comparison of measured and predicted effective areas (EA) at $\sim$ 4.6 keV. Effective area 1-$\sigma$ uncertainties are $\sim$ $\pm$3\% for predicted values and $\pm$4\% for measured values. Difference column is (measured-predicted)/average value, expressed as a percentage. The mean difference value over the data set is -3.0\%, with a standard deviation of 3.1\%}.}
\centering
\begin{tabular}{| c c c c c |}
\hline
\multicolumn{2}{|c}{\textbf{MMA}} & \textbf{Predicted} & \textbf{Measured} & \textbf{Difference}\\
\multicolumn{2}{|c}{\textbf{(arcmin)}} & \textbf{Telescope} & \textbf{Telescope} & \textbf{$\%$}\\
\textbf{Pan (X)} & \textbf{Tip (Y)} & \multicolumn{1}{r}{\textbf{EA (cm\textsuperscript{2})}} &  \multicolumn{1}{r}{\textbf{EA (cm\textsuperscript{2})}} & \\
\hline
0 & \multicolumn{1}{c}{0} & 7.5 & 7.4 & -1.3 \\
\hline
0 & \multicolumn{1}{c}{5.0} & 5.8 & 5.6 & -2.6 \\
\hline
0 & \multicolumn{1}{c}{-5.0} & 5.8 & 5.5 & -4.4 \\
\hline
5.0 & \multicolumn{1}{c}{0} & 5.7 & 5.6 & -2.2 \\
\hline
-5.0 & \multicolumn{1}{c}{0} & 5.7 & 5.6 & -2.3 \\
\hline
4.95 & \multicolumn{1}{c}{4.95} & 4.9 & 4.7 & -5.8 \\
\hline
4.95 & \multicolumn{1}{c}{-4.95} & 4.9 & 4.7 & -4.6 \\
\hline
-4.95 & \multicolumn{1}{c}{4.95} & 4.9 & 4.5 & -7.9 \\
\hline
-4.95 & \multicolumn{1}{c}{-4.95} & 5.0 & 4.5 & -8.9 \\
\hline
0 & \multicolumn{1}{c}{3.0} & 6.6 & 6.5 & -1.6 \\
\hline
0 & \multicolumn{1}{c}{-3.0} & 6.6 & 6.2 & -5.3 \\
\hline
3.0 & \multicolumn{1}{c}{0} & 6.5 & 6.7 & 2.3\\
\hline
-3.0 & \multicolumn{1}{c}{0} & 6.5 & 6.7 & 1.8 \\
\hline
2.83 & \multicolumn{1}{c}{2.83} & 6.1 & 5.9 & -3.1 \\
\hline
2.83 & \multicolumn{1}{c}{-2.83} & 6.1 & 6.2 & 0.9 \\
\hline
-2.83 & \multicolumn{1}{c}{2.83} & 6.1 & 6.0 & -2.4 \\
\hline
-2.83 & \multicolumn{1}{c}{-2.83} & 6.1 & 5.9 & -2.7 \\
\hline
\end{tabular}
\label{tab:EffArea4.6keV}
\end{table}
\begin{table}[htp]
\caption{\textit{Comparison of measured and predicted effective areas (EA) at $\sim$ 6.6 keV. Effective area 1-$\sigma$ uncertainties are $\sim$ $\pm$3\% for predicted values and $\pm$2\% for measured values. Difference column is (measured-predicted)/average value, expressed as a percentage. The mean difference value over the data set is 1.5\%, with a standard deviation of 2.7\% }.}
\centering
\begin{tabular}{| c c c c c |}
\hline
\multicolumn{2}{|c}{\textbf{MMA}} & \textbf{Predicted} & \textbf{Measured} & \textbf{Difference}\\
\multicolumn{2}{|c}{\textbf{(arcmin)}} & \textbf{Telescope} & \textbf{Telescope} & \textbf{$\%$}\\
\textbf{Pan (X)} & \textbf{Tip (Y)} & \multicolumn{1}{r}{\textbf{EA (cm\textsuperscript{2}) }} &  \multicolumn{1}{r}{\textbf{EA (cm\textsuperscript{2})}} & \\
\hline
0 & \multicolumn{1}{c}{0} & 2.68 & 2.82 & 4.9 \\
\hline
0 & \multicolumn{1}{c}{5.0} & 1.70 & 1.71 & 0.1 \\
\hline
0 & \multicolumn{1}{c}{-5.0} & 1.70 & 1.74 & 2.0 \\
\hline
5.0 & \multicolumn{1}{c}{0} & 1.76 & 1.75 & -0.6\\
\hline
-5.0 & \multicolumn{1}{c}{0} & 1.76 & 1.76 & -0.3\\
\hline
4.95 & \multicolumn{1}{c}{4.95} & 1.38 & 1.36 & -1.3\\
\hline
4.95 & \multicolumn{1}{c}{-4.95} & 1.34 & 1.31 & -2.4\\
\hline
-4.95 & \multicolumn{1}{c}{4.95} & 1.34 & 1.30 & -3.2\\
\hline
-4.95 & \multicolumn{1}{c}{-4.95} & 1.38 & 1.35 & -1.8\\
\hline
0 & \multicolumn{1}{c}{3.0} & 2.15 & 2.27 & 5.4\\
\hline
0 & \multicolumn{1}{c}{-3.0} & 2.15 & 2.26 & 4.9\\
\hline
3.0 & \multicolumn{1}{c}{0} & 2.18 & 2.28 & 4.6\\
\hline
-3.0 & \multicolumn{1}{c}{0} & 2.18 & 2.23 & 2.4\\
\hline
2.83 & \multicolumn{1}{c}{2.83} & 1.92 & 1.97 & 2.3\\
\hline
2.83 & \multicolumn{1}{c}{-2.83} & 1.91 & 1.97 & 2.9\\
\hline
-2.83 & \multicolumn{1}{c}{2.83} & 1.91 & 1.97 & 2.9\\
\hline
-2.83 & \multicolumn{1}{c}{-2.83} & 1.92 & 1.98 & 3.0\\
\hline
\end{tabular}
\label{tab:EffArea6.6keV}
\end{table}
\clearpage

\subsection{Modulation Factor}\label{sec:MF}
\subsubsection{Calibration Requirements}
The modulation factor ($\mu$) shall be measured to an accuracy equal to or better than  1$\%$ of its value for at least 2 energies in the range
1.5--8.0 keV and with at least 2 dither patterns at each energy.

\subsubsection{Measurements}
The measurement of the telescope modulation factor started with characterization of the width of the polarized beam for each source,
generated by diffracting X-rays from a nominal point source off specific crystals at precise angles. This characterization
was necessary as the width of the polarized beam in the dispersion direction was significantly less than the width of the MMA, thus
the beam must be stepped through multiple positions to ensure uniform coverage of the optic. The beam characterization is performed,
for each polarized source, by driving the MMA to its out-of-focus position (+ 50 mm from true focus) and acquiring an image on the
focal plane CCD detector. Analysis of this ring image showed the extent of coverage by the beam, and allowed calculation of the number
of discrete steps (achieved typically by 0.035$^\circ$ rotations of the crystal) needed to give uniform MMA coverage. These images at discrete steps
were summed and the resulting ring image analyzed for uniformity. When acceptable (greater than 90$\%$ uniformity across the MMA active surface), these parameters were recorded for use when
making the modulation factor measurement. This process was repeated for all sources used.

Telescope modulation factors were measured for three source + crystal combinations (see Table \ref{tab:Crystals}), with source energies and crystal
lattice spacing chosen to achieve Bragg angles close to 45$^\circ$. Consequently, the resulting X-ray beams are nearly totally (linearly) polarized.
For all measurements of modulation factors, dithering of the MMA was used to simulate the dithering planned for use on orbit.
This dithering also approximately aligns with that used for detector calibrations at INAF-IAPS. Two dithering values were used for the telescope calibration: 1.78 mm radius
and 3.57 mm radius, implemented by lateral motions of the MMA-supporting hexapod. The dithering pattern was a Lissajous-type figure,
designed to provide near uniform coverage of the dither area.
The required detector counts to achieve the required 1$\%$ accuracy in measured modulation factor are given in Table \ref{tab:ModFactDetectorCounts}, along with the
actual counts accrued during the measurement.

\begin{table}[ht!]
\centering
\caption{\textit{Modulation factor measurement dither and count parameters}.}
\label{tab:ModFactDetectorCounts}
\begin{tabular}{| c c c c |}
\hline
\textbf{Source} & \textbf{Dither Radius (mm)} & \textbf{Counts Required} & \textbf{Counts Accrued} \\
\hline
\text{Rh (L)} & 1.78 & 3.5x10\textsuperscript{5} & 1.2x10\textsuperscript{6} \\
\hline
\text{Ti (K)} & 1.78 & 1.2x10\textsuperscript{5} & 2.9x10\textsuperscript{5} \\
\hline
\text{Fe (K)} & 1.78 & 8.6x10\textsuperscript{4} & 1.9x10\textsuperscript{5} \\
\hline
  &   &   &   \\
\hline
\text{Rh (L)} & 3.57 & 1.1x10\textsuperscript{6} & 1.6x10\textsuperscript{6} \\
\hline
\text{Ti (K)} & 3.57 & 4.0x10\textsuperscript{5} & 7.1x10\textsuperscript{5} \\
\hline
\text{Fe (K)} & 3.57 & 2.9x10\textsuperscript{5} & 5.8x10\textsuperscript{5} \\
\hline

\end{tabular}

\end{table}

\subsubsection{Results}\label{sec:MFRes}
\begin{figure}[ht!]
    \centering
    \includegraphics[width=0.6\linewidth]{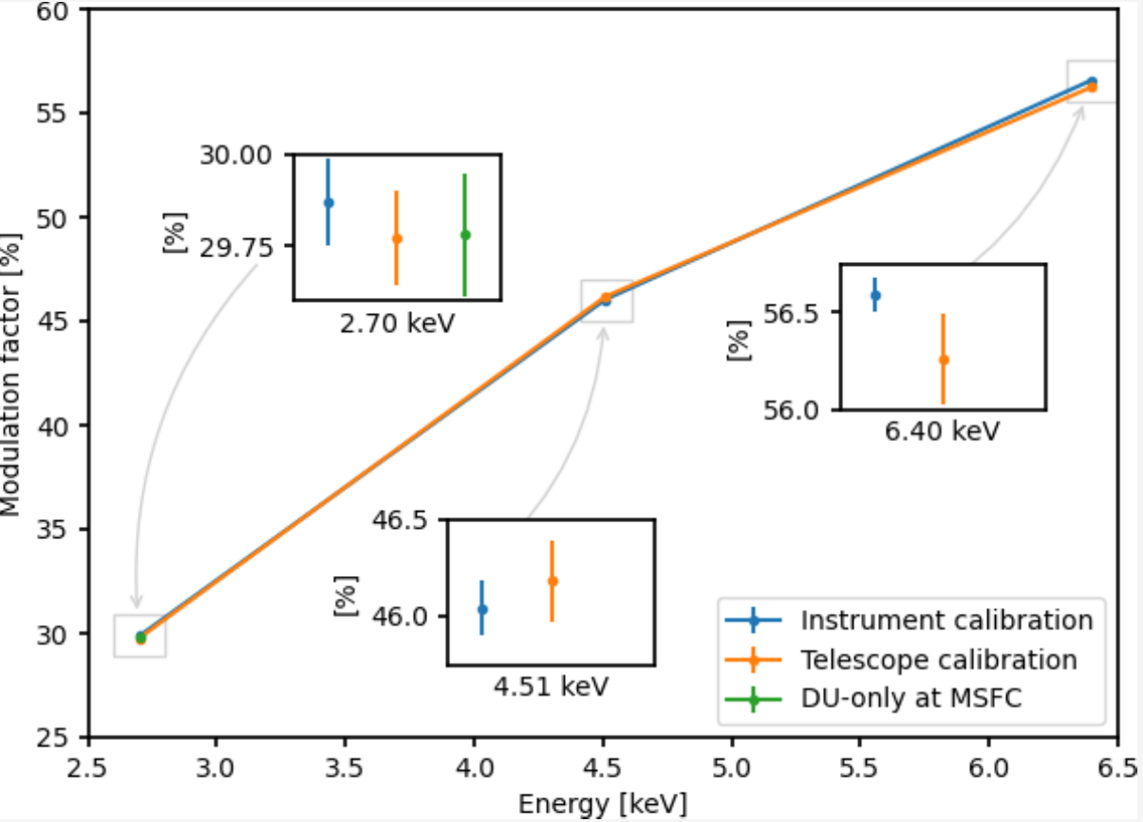}
    \caption{\textit{The summary of the modulation factor calibration at MSFC and at INAF-IAPS  (3.57 mm dither) showing good
    agreement}.}
    \label{fig:ModFactRecap}
\end{figure}

In order to accurately assess the modulation factor it was necessary to remove any spurious modulation measured for each energy during the separate
calibration of this detector. Then the DU data were processed using a standard moments analysis with a 20$\%$ cut on tracks, based on track ovality,
and an energy cut of $\pm 3\sigma$ around all line energies (see \cite{Baldini2021} for a description of detector data analysis). The measured
modulation factors for the telescope are then compared to those measured for the DU alone during detector calibration by IAPS at INAF-IAPS. The results
are shown in Table \ref{tab:ModFactMeasured}, together with the 1-$\sigma$ measurement errors, and in Figure \ref{fig:ModFactRecap}.

\begin{table}[t]
\caption{\textit{Measured telescope and DU modulation factors. Difference column is detector value-telescope value}. }. 
\label{tab:ModFactMeasured}
\begin{center}
\end{center}
\begin{tabular}{| C C C C C|}
\hline
\textbf{Source} & \textbf{Dither Radius} & \textbf{Telescope Modulation Factor} & \textbf{Detector Modulation Factor} & \textbf{Difference}\\
\textbf{(Energy)} & (\textbf{mm}) & \textbf{DU-FM1} & \textbf{DU-FM1 @ IAPS} & \\
& & \textbf{\%} & \textbf{\%} & \\
\hline
\text{Rh L (2.7 keV)} & 1.78 & 29.69 $\pm$ 0.20 & 29.87 $\pm$ 0.12 & 0.18 \\
\hline
\text{Ti K (4.5 keV)} & 1.78 & 45.43 $\pm$ 0.33 & 46.04 $\pm$ 0.14 & 0.61\\
\hline
\text{Fe K (6.4 keV)} & 1.78 & 57.08 $\pm$ 0.42 & 56.59 $\pm$ 0.09 & -0.49\\
\hline
  &   &   &   &  \\
\hline
\text{Rh L (2.7 keV)} & 3.57 & 29.77 $\pm$ 0.13 & 29.87 $\pm$ 0.13 & 0.10\\
\hline
\text{Ti K (4.5 keV)} & 3.57 & 46.18 $\pm$ 0.21 & 46.04 $\pm$  0.14 & -0.14\\
\hline
\text{Fe K (6.4 keV)} & 3.57 & 56.26 $\pm$ 0.23 & 56.59 $\pm$ 0.09 & 0.33\\
\hline

\end{tabular}

\end{table}

As an additional check that nothing had changed between the original detector calibration at INAF-IAPS and the telescope calibration much later at MSFC,
an additional set of data was taken at the Stray Light Test Facility, moving the detector unit so that it viewed the polarized source directly, rather
than through the MMA. Because the DU is rate limited to $\sim$ 150 c/s it was necessary to electronically mask the DU to accept only events within a
3-mm-radius circle to approximate the dither area with the MMA. In this way the full 150 c/s could be applied to the central region and the desired number
of counts could be obtained in a reasonable timescale. These results are shown in Table \ref{tab:ModFact@MSFC}.

\begin{table}[ht!]
\centering
\caption{\textit{Measurements of modulation factors for DU-FM1 at MSFC, with and without optics, compared with those taken at INAF-IAPS}.}
\label{tab:ModFact@MSFC}
\begin{tabular}{| c c c c |}
\hline
\textbf{Source / Energy} & \textbf{Modulation Factor} & \textbf{Modulation Factor} & \textbf{Modulation Factor}\\
& \textbf{DU-FM1 at MSFC} & \textbf{DU-FM1 + MMA4} & \textbf{DU-FM1 @ IAPS}\\
\hline
\text{Rh L (2.7 keV)} & 29.78 $\pm$ 0.17 \% & 29.77 $\pm$ 0.13 \% & 29.87 $\pm$ 0.12 \% \\
\hline
\end{tabular}
\end{table}

\subsubsection{Discussion}\label{MF Discussion}
The above data show that the requirement for the number of energies and for statistical precision of modulation factor measurements were met.
Table \ref{tab:ModFactMeasured} and Table \ref{tab:ModFact@MSFC} show that modulation factors measured with the telescope are equal to those measured
with the DU alone within the statistical uncertainties. This shows that for the modulation factor the telescope response can be derived from the DU calibration data alone, and that
there are no discernible MMA-induced effects.

\subsection{Spurious Modulation}\label{sec:SM}
\subsubsection{Calibration requirement}\label{sec:SMCalReq}
The spurious-modulation amplitude shall be measured to an accuracy equal to or better than 0.1$\%$ absolute value, for at least 2 energies in the range 1.5--8 keV and
with at least 1 dither pattern at each energy.

\subsubsection{Measurements}\label{sec:SMMeas}
This measurement utilizes low-flux non-polarized X-ray sources to evaluate low-level modulation inherent in the response of the detector. The collected counts required to do this at the required absolute accuracy (0.1\%) are given in Table \ref{tab:SpuriousModul}, along with the total counts accrued during the measurement.

The X-ray sources, however, can have a small degree of residual polarization, due to the way the X rays are generated. In electron-impact X-ray sources, those sources with end windows, where the emitted X-rays are parallel to the electron direction, exhibit no polarization whereas those sources whose X-ray emission is perpendicular to the electron beam show intrinsic polarization \citep{Ratheesh2023a}. As the sources used in the IXPE telescope calibration are of the latter kind, this source polarization must be removed from the data. This is accomplished, for each X-ray source, by splitting the measurement in two and rotating the x-ray head by 90 degrees for the second half of the accumulation. Adding together the two halves then cancels out the effects of any intrinsic polarization of the source.

As with the modulation factor measurements, the optic was dithered with a radius of 3.57 mm. However, for the aluminum source the hexapod dither had become unreliable (a lubrication issue) and so the image was spread by simply defocusing the optic by 50 mm to give a suitably sized focal region.

\begin{table}[ht!]
\centering
\caption{\textit{Spurious modulation measurements dither and count parameters}.} 

\label{tab:SpuriousModul}
\center
\begin{tabular}{| c c c c |}
\hline
Source / Energy & Dither Radius (mm) & Counts Required & Counts Accrued \\
\hline
Mo L (2.3 keV) & 3.57 & 3.0 x 10\textsuperscript{6} & 3.3x10\textsuperscript{6} \\
\hline
Ti K (4.6 keV) & 3.57 & 3.0 x 10\textsuperscript{6} & 3.2x10\textsuperscript{6} \\
\hline
Fe K (6.6 keV) & 3.57 & 3.0 x 10\textsuperscript{6} & $^{a} $2.6x10\textsuperscript{6} \\
\hline
Al K (1.5 keV) & Not dithered, but defocused & 3.0 x 10\textsuperscript{6} & 3.1x10\textsuperscript{6} \\
\hline
\end{tabular}    
\footnote{Note that the required counts for Fe were not precisely achieved due to problems with the dither near the end of the run. Nevertheless, the calibration requirement of ‘at least 2 energies’ was met.}
\end{table}

\subsubsection{Results}\label{sec:SMResults}
Data were processed without applying any track cuts. Table \ref{tab:SpurModNet} shows the spurious modulation results for the four sources with the intrinsic source
modulation removed as described above. For the aluminum source the data were taken from a defocused ring image (no dithering) and analyzed within a 1.5-mm radius. Measured amplitudes of the intrinsic source modulations, which reflect an intrinsic polarization of the X-ray sources, is given
in Table \ref{tab:ModResTubes} for reference.

\begin{table}[ht!]
\centering
\caption{\textit{Telescope spurious modulation measurements after X-ray source modulation removal}.}
\label{tab:SpurModNet}
\begin{tabular}{| c c c |}
\hline
\textbf{Source / Energy} &\textbf{ Mod (\%)} & \textbf{Mod-error} \\
& & \textbf{($\%$)} \\
\hline
Mo L (2.3 keV) & 0.373 & 0.084 \\
\hline
Ti K (4.6 keV) & 0.056 & 0.097 \\
\hline
Fe K (6.6 keV) & 0.24 & 0.11 \\
\hline
Al K (1.5 keV) & 0.5 & 0.11 \\
\hline
\end{tabular}
\end{table}

\begin{table}[ht!]
\centering
\caption{\textit{Measured modulations from the calibration electron-impact sources}. }
\label{tab:ModResTubes}
\begin{tabular}{| c c c |}
\hline
\textbf{Source / Energy} & \textbf{Mod} & \textbf{Mod-error} \\
& (\%) & (\%) \\ 
\hline
\text{Mo L (2.3 keV)} & 0.254 & 0.085 \\
\hline
\text{Ti K (4.5 keV)} & 0.503 & 0.097 \\
\hline
\text{Fe K (6.4 keV)} & 0.28 & 0.11 \\
\hline
\text{Al K (1.5 keV)} & 0.441 & 0.089 \\
\hline

\end{tabular}

\end{table}

\begin{figure}
\centering
    \includegraphics[width=0.5\linewidth]{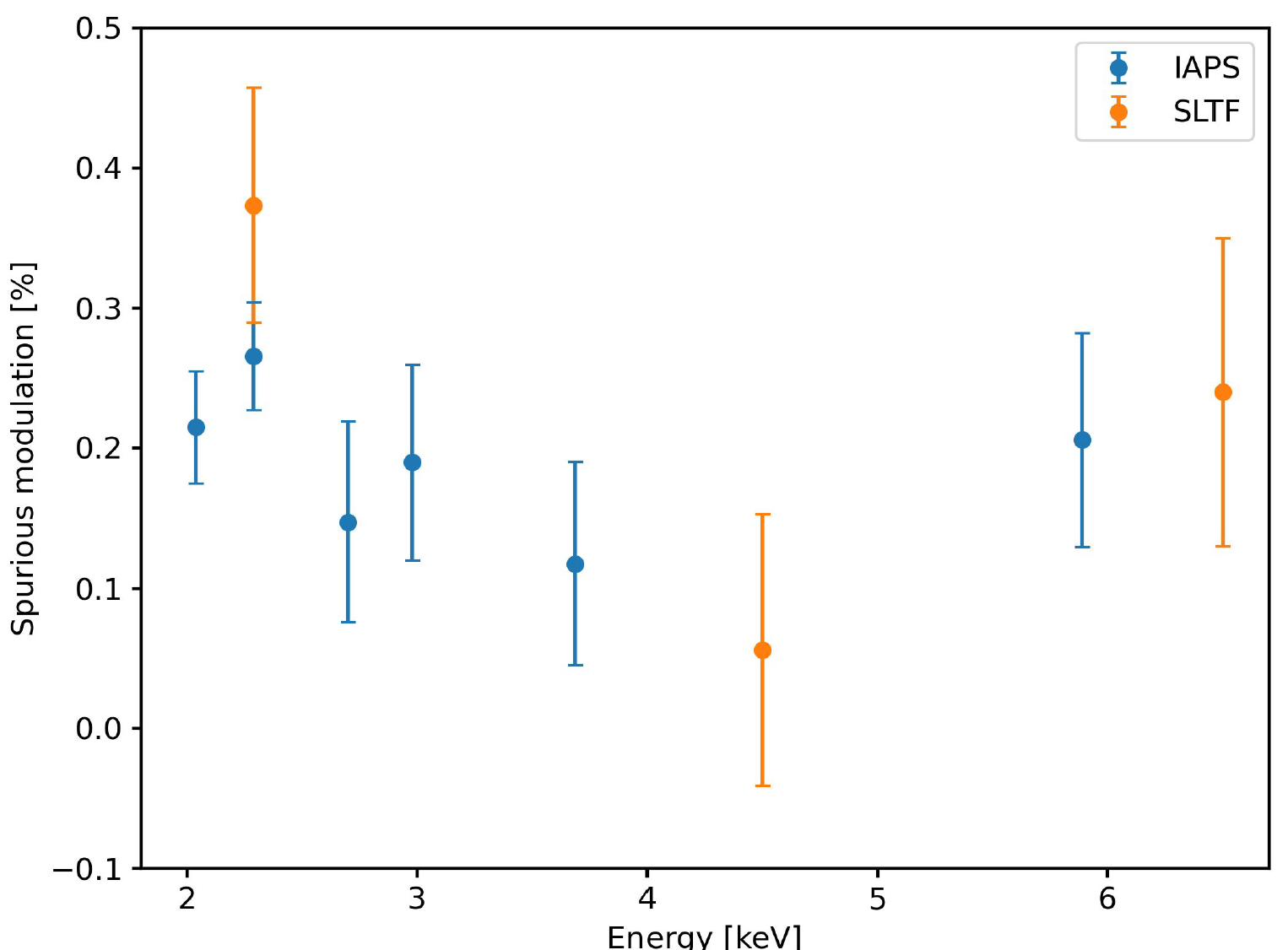}
\caption{\textit{Spurious modulation measurements at MSFC compared with data taken at INAF-IAPS on a flat field illumination of 3.57 mm radius}}
\label{fig:SMMSFCIAPS}
\end{figure}

Figure \ref{fig:SMMSFCIAPS} compares the spurious modulation measured at 3 energies during telescope calibration at MSFC (orange) with measurements of the detector (DU-FM1) at IAPS (blue).
Although the energies are not all the same, it is evident that the two data sets are in good agreement (1-$\sigma$ errors shown).

An additional set of measurements were performed at SLTF at 2.3 keV, where the spurious modulation was measured for the telescope and then for DU-FM1 alone, after moving from behind the optic so that it was directly illuminated by the source. For the latter measurement, an electronic mask was applied to the DU so that only events within the dither radius were registered (as for the modulation factor measurements). Note that this extraction radius was 10\% smaller than for the data in Figure \ref{fig:SMMSFCIAPS}, and as the spurious modulation is position dependent the small difference between the data at 2.3 keV in Figure 7 and Table 15 is to be expected. 

\begin{table}
\centering
\caption{\textit{Spurious modulation measurement results at 2.3 keV with and without MMA-4 and and an extraction region of 3.3 mm radius at MSFC.}}
\label{tab:SpurMod2.3keV}
\begin{tabular}{| l l l l |}
\hline
Source /Energy & Mod (\%) & Mod-error (\%) & MMA present \\
\hline
Mo L (2.3 keV) & 0.531 & 0.089  & Yes \\
\hline
Mo L (2.3 keV) & 0.41 & 0.10 & No \\
\hline

\end{tabular}

\end{table}

\subsubsection{Discussion}\label{sec:SMDisc}
The above data show first that the requirement for spurious modulation measurement was met for the telescope calibration: At least two energies were used, and a statistical precision of $\le$ 0.1 \% absolute (see Mod-error column of Table \ref{tab:SpurModNet} ) was achieved for two energies, and within rounding errors, for an additional 2 energies.

The data also show that the MMA has no effect on spurious modulation. Figure \ref{fig:SMMSFCIAPS} compares data taken during telescope calibration with data taken with the same DU-FM1 during detector calibration at INAF-IAPS. These spurious modulation data are in good agreement. Table \ref{tab:SpurMod2.3keV} shows measurements at 2.3 keV both at the focus of MMA4 and then in the direct beam at the SLTF without the optic. The spurious modulation (Mod \%) for the two measurements are within 1-$\sigma$ of each other (see Mod-error column) and so are statistically identical. As with the modulation factor, the spurious modulation is unchanged by the presence of the optic.

\subsection{Half Power Diameter}\label{sec:HPD}
\subsubsection{Calibration Requirement}\label{sec:HPDCalReq}
The half-power diameter (HPD) of each MMA-DU combination tested shall be measured to an accuracy equal to or better than 3\% of its value, on axis and at least 16 off-axis positions, for at least 3 energies between 1.5 and 8 keV.

\subsubsection{Measurements}\label{sec:HPDMeas}
As for the effective area measurements, DU-FM1 was decentered and a best focus position was obtained, minimizing the measured angular resolution. The required HPD data and effective area data were collected simultaneously stepping through the optic angles and appropriate detector positions that were given in Table \ref{tab:OpticsPosition}.

The half-power diameter of each MMA was measured using the Andor CCD camera at the focus of the optic, with filtered X-ray sources producing lines at nominally 2.3 keV (MoL), 4.6 keV (TiK) and 6.6 keV (FeK) along with some continuum emission. The detector region used for this measurement was chosen to be the same as that used for the effective area, i.e. a circle of diameter 8 mm, which is approximately 16 times the MMA HPD. Flat fields, taken immediately before and after the measurements, were used to subtract offsets and noise contributions in individual pixels. The diameter containing half the flux within the measurement region was then determined and converted to an angle using the measured image distance ($\sim$4.17m) for the 100-m object distance at the SLTF.

For the telescope, the HPD was measured at the same energies as was the MMA, using the same 8-mm-diameter circular region on the DU. As with the MMA, the diameter containing half the flux within this region was calculated and converted to an angle, again using the same measured focal distance (image distance). 

It is expected that the telescope and MMA HPDs will differ slightly. This is because the DU adds additional blurring to the image due to its finite spatial resolution (limited by the ability to determine the beginning of a track) and by the finite depth of the detector (10 mm), which adds additional defocusing of the image due to gas transparency and the cone angle of X rays focused by the mirror assembly. However, these two components are small compared to the native MMA resolution.

The detector spatial resolution was measured during calibration at INAF-IAPS.  An additional factor has been added to these reported spatial resolutions to account for an 0.8$^\circ$ misalignment (estimated after the calibration) of the DU with respect to the X-ray axis during the telescope calibration. 

Defocusing effects, due to finite detector gas depth, were calculated via Monte-Carlo simulations. An initial value for this contribution, modeled for a perfect optic, to be added in quadrature to the native optic resolution was found to be 8.1 arcsec. However, using measured MMA figure and circularity errors in the simulation increased this effect to 9.4 arcsec, independent of energy as the X-ray mean-free-path is larger than the detector depth at all tested energies. As a check on this Monte-Carlo approach, results were also obtained by taking a series of CCD images over a +/- 5-mm region centered on the optimum focal distance, to span the region covered by the thickness of the detector gas cell. Summing these CCD images effectively blurs the image in a way similar to that due to the gas depth in the DU. These measurements agreed well with the Monte-Carlo simulations (within one arcsecond).

\subsubsection{Results}\label{sec:HPDRes}
Table \ref{tab:HPDOnAxis} shows the results of the on-axis HPD comparison. MMA HPD represents the angular resolution of the MMA measured with the CCD camera. Detector spatial resolution and defocusing effects, converted to HPD in arcsec, are summed in quadrature with the native MMA resolution to give the predicted telescope angular resolution. The next column shows the measured telescope HPD obtained during telescope calibration. Estimated uncertainties are $\sim$2$\%$ (1-$\sigma$) for the predicted HPD measurements and $\sim$0.5-1$\%$ for the measured value. As can be seen from the table, the predicted and measured telescope HPDs are in reasonable agreement given measurement uncertainties. 

\begin{table}[ht!]
\centering
\caption{\textit{Comparison of predicted and measured on-axis telescope angular resolution.}}
\label{tab:HPDOnAxis}
\begin{tabular}{| C C C C C C C |}
\hline
\textbf{X-ray Tube} & \textbf{MMA} & \textbf{Detector} & \textbf{Detector} & \textbf{Predicted} & \textbf{Measured} & \textbf{Difference}\\
\textbf{(Line)} & \textbf{HPD} & \textbf{Spatial} & \textbf{Defocusing} & \textbf{Telescope} & \textbf{Telescope} & \textbf{\%}\\
 & \textbf{(arcsec)} & \textbf{Resolution} & \textbf{Effects} & \textbf{HPD} & \textbf{HPD} & \\
 &  & \textbf{(arcsec)} & \textbf{(arcsec)} & \textbf{(arcsec)} & \textbf{(arcsec)} & \\
\hline
\text{2.3 keV} & 20.0 & 5.6 & 9.4 & 22.8 & 22.2 & -2.7 \\
\hline
\text{4.6 keV} & 20.8 & 6.3 & 9.4 & 23.7 & 23.8 & 0.6\\
\hline
\text{6.6 keV} & 20.1 & 7.4 & 9.4 & 23.4 & 24.1 & 2.8\\
\hline
\end{tabular}
\end{table}

This same process was repeated, at each energy, for all off-axis angles (Tables \ref{tab:HPD2.3keV}--\ref{tab:HPD6.6keV}).

\begin{table}

\caption{\textit{Comparison of predicted and measured telescope angular resolution at 2.3 keV
HPD 1-$\sigma$ uncertainties are $\sim$ $\pm$ 2$\%$ for predicted values and $\pm$ 0.5$\%$ for measured values. Difference column is (measured-predicted)/average value, expressed as a percentage.The mean difference value over the data set is -2.1\%, with a standard deviation of 2.6\% }. 
}
\label{tab:HPD2.3keV}
\centering
\begin{tabular}{| c c c c c |}
\hline
\multicolumn{2}{|c}{\textbf{MMA}} & \textbf{Predicted} & \textbf{Measured} & \textbf{Difference}\\
\multicolumn{2}{|c}{\textbf{(arcmin)}} & \textbf{Telescope HPD} & \textbf{Telescope HPD} & \textbf{\%}\\
\textbf{Pan (X)} & \textbf{Tip (Y)} & \textbf{(arcsec)} &  \textbf{(arcsec)} & \\
\hline
0 & \multicolumn{1}{c}{0} & 22.8 & 22.2 & -2.7\\
\hline
0 & \multicolumn{1}{c}{5.0} & 24.2 & 23.5 & -3.0\\
\hline
0 & \multicolumn{1}{c}{-5.0} & 24.1 & 23.8 & -1.5\\
\hline
5.0 & \multicolumn{1}{c}{0} & 24.1 & 24.9 & 3.1\\
\hline
-5.0 & \multicolumn{1}{c}{0} & 24.2 & 23.4 & -3.6\\
\hline
4.95 & \multicolumn{1}{c}{4.95} & 24.9 & 24.7 & -1.0\\
\hline
4.95 & \multicolumn{1}{c}{-4.95} & 25.7 & 26.1 & 1.5\\
\hline
-4.95 & \multicolumn{1}{c}{4.95} & 25.9 & 23.8 & -8.3\\
\hline
-4.95 & \multicolumn{1}{c}{-4.95} & 25.7 & 24.2 & -5.8\\
\hline
0 & \multicolumn{1}{c}{3.0} & 23.5 & 23.1 & -1.9\\
\hline
0 & \multicolumn{1}{c}{-3.0} & 23.5 & 23.1 & -1.9\\
\hline
3.0 & \multicolumn{1}{c}{0} & 23.9 & 23.7 & -0.8\\
\hline
-3.0 & \multicolumn{1}{c}{0} & 23.6 & 22.9 & -2.9\\
\hline
2.83 & \multicolumn{1}{c}{2.83} & 23.9 & 24.0 & 0.3\\
\hline
2.83 & \multicolumn{1}{c}{-2.83} & 24.2 & 24.0 & -0.9\\
\hline
-2.83 & \multicolumn{1}{c}{2.83} & 23.9 & 23.0 & -3.5\\
\hline
-2.83 & \multicolumn{1}{c}{-2.83} & 24.1 & 23.4 & -2.7\\
\hline
\end{tabular}

\end{table}

\begin{table}[ht!]
\centering
\caption{\textit{Comparison of predicted and measured telescope angular resolution at 4.6 keV
HPD 1-$\sigma$ uncertainties are $\sim$ $\pm$2\% for predicted values and $\pm$ 1\% for measured values.Difference column is (measured-predicted)/average value, expressed as a percentage. The mean difference value over the data set is 1.4\%, with a standard deviation of 2.3\%}. 
}
\label{tab:HPD4.6keV}
\begin{tabular}{| c c c c c |}
\hline
\multicolumn{2}{|c}{\textbf{MMA}} & \textbf{Predicted} & \textbf{Measured} & \textbf{Difference}\\
\multicolumn{2}{|c}{\textbf{(arcmin)}} & \textbf{Telescope HPD} & \textbf{Telescope HPD} & \textbf{\%}\\
\textbf{Pan (X)} & \textbf{Tip (Y)} & \textbf{(arcsec)} &  \textbf{(arcsec)} & \\
\hline
0 & \multicolumn{1}{c}{0} & 23.7 & 23.8 & 0.6\\
\hline
0 & \multicolumn{1}{c}{5.0} & 25.4 & 25.6 & 0.7\\
\hline
0 & \multicolumn{1}{c}{-5.0} & 25.6 & 25.5 & -0.5\\
\hline
5.0 & \multicolumn{1}{c}{0} & 25.1 & 26.2 & 4.3\\
\hline
-5.0 & \multicolumn{1}{c}{0} & 25.5 & 25.4 & -0.1\\
\hline
4.95 & \multicolumn{1}{c}{4.95} & 26.2 & 26.4 & 0.7\\
\hline
4.95 & \multicolumn{1}{c}{-4.95} & 27.1 & 27.9 & 2.9\\
\hline
-4.95 & \multicolumn{1}{c}{4.95} & 26.8 & 25.9 & -3.6\\
\hline
-4.95 & \multicolumn{1}{c}{-4.95} & 26.8 & 26.4 & -1.5\\
\hline
0 & \multicolumn{1}{c}{3.0} & 24.4 & 25.0 & 2.5\\
\hline
0 & \multicolumn{1}{c}{-3.0} & 24.4 & 25.0 & 2.6\\
\hline
3.0 & \multicolumn{1}{c}{0} & 24.5 & 25.7 & 5.0\\
\hline
-3.0 & \multicolumn{1}{c}{0} & 24.8 & 24.9 & 0.6\\
\hline
2.83 & \multicolumn{1}{c}{2.83} & 24.9 & 25.8 & 3.4\\
\hline
2.83 & \multicolumn{1}{c}{-2.83} & 24.9 & 26.0 & 4.3\\
\hline
-2.83 & \multicolumn{1}{c}{2.83} & 24.8 & 24.8 & -0.2\\
\hline
-2.83 & \multicolumn{1}{c}{-2.83} & 25.1 & 25.4 & 1.3\\
\hline

\end{tabular}
\end{table}

\begin{figure}[htb!] \centering
\subfigure[\label{subFig:PSFMAPMo}]{\includegraphics[height=7.0
cm]{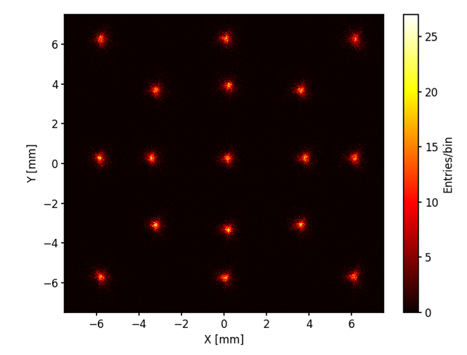}}
\subfigure[\label{subFig:PSFModelMo}]{\includegraphics[height =
6.6 cm]{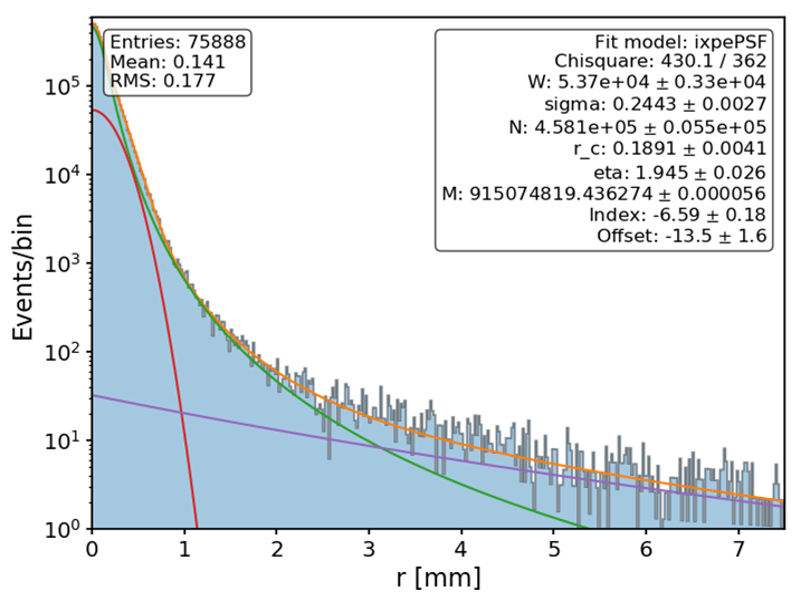}} \label{subFig:MoX-rayTubePSF} \caption{\textit{ ({\bf a}) Source images obtained at an average energy of 2.3 keV demonstrates good off-axis performance compared to on-axis performance.}
\textit{({\bf b}) The on-axis PSF of the telescope is modeled as the sum of a Gaussian core and a King function, along with a power-law component that dominates the outer regions of the field of view}.}
\end{figure}

\begin{figure}[htb!] \centering
\subfigure[\label{subFig:PSFMAPTi}]{\includegraphics[height=7.0
cm]{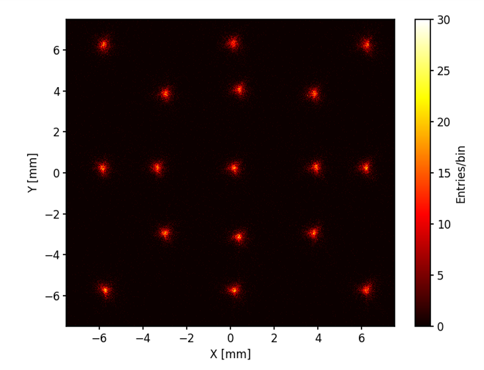}}
\subfigure[\label{subFig:PSFModelTi}]{\includegraphics[height =
6.6 cm]{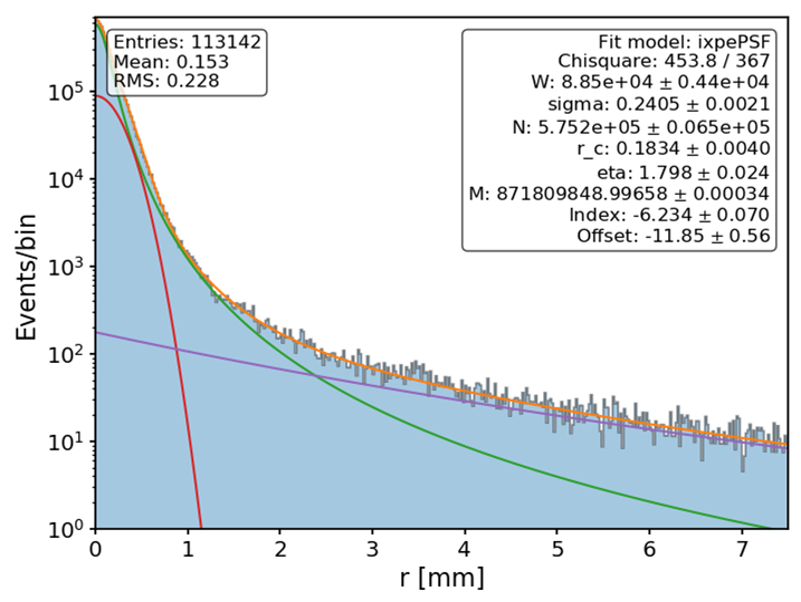}}  \caption{\textit{({\bf a}) Source images obtained at an average energy of 4.6 keV. }
\textit{({\bf b}) The on-axis PSF of the telescope is modeled as the sum of a Gaussian core and a King function, along with a power-law component that dominates the outer regions of the field of view}.}
\end{figure}

\begin{figure}[htb!] \centering
\subfigure[\label{subFig:PSFMAPFe}]{\includegraphics[height=7.0
cm]{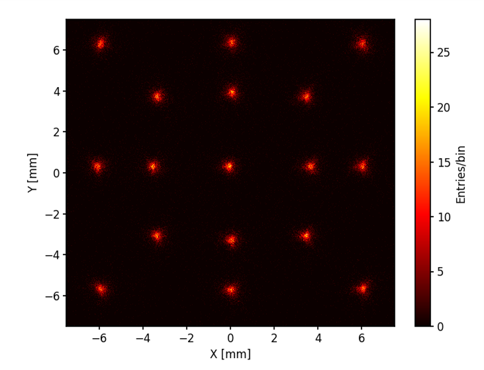}}
\subfigure[\label{subFig:PSFModelFe}]{\includegraphics[height =
6.6 cm]{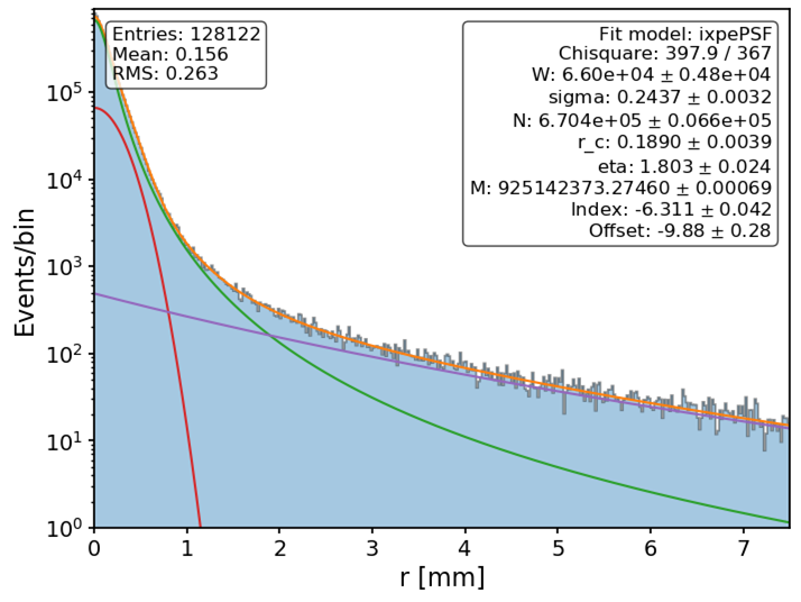}} \label{subFig:FeX-rayTubePSF} \caption{\textit{({\bf a}) Source images obtained at an average energy of 6.6 keV.}
\textit{({\bf b}) The on-axis PSF of the telescope is modeled as the sum of a Gaussian core with a King function, along with a power-law component that dominates the outer regions of the field of view}.}
\end{figure}

\begin{table}[ht!]
\centering

\caption{\textit{Comparison of predicted and measured telescope angular resolution at 6.6 keV HPD 1-$\sigma$ uncertainties are $\sim$ $\pm$2\% for predicted values and $\pm$ 1\% for measured values. Difference column is (measured-predicted)/average value, expressed as a percentage. The mean difference value over the data set is 5.7\%, with a standard deviation of 2.6\%}.
}
\label{tab:HPD6.6keV}
\begin{tabular}{| c c c c c |}
\hline
\multicolumn{2}{|c}{\textbf{MMA}} & \textbf{Predicted} & \textbf{Measured} & \textbf{Difference}\\
\multicolumn{2}{|c}{\textbf{(arcmin)}} & \textbf{Telescope HPD} & \textbf{Telescope HPD} & \textbf{\%}\\
\textbf{Pan (X)} & \textbf{Tip (Y)} & \textbf{(arcsec)} &  \textbf{(arcsec)} & \\
\hline
0 & \multicolumn{1}{c}{0} & 23.4 & 24.1 & 2.8\\
\hline
0 & \multicolumn{1}{c}{5.0} & 25.2 & 26.3 & 4.1\\
\hline
0 & \multicolumn{1}{c}{-5.0} & 25.9 & 27.9 & 7.7\\
\hline
5.0 & \multicolumn{1}{c}{0} & 25.2 & 27.2 & 7.8\\
\hline
-5.0 & \multicolumn{1}{c}{0} & 25.9 & 26.6 & 3.0\\
\hline
4.95 & \multicolumn{1}{c}{4.95} & 25.9 & 26.9 & 3.9\\
\hline
4.95 & \multicolumn{1}{c}{-4.95} & 26.7 & 28.7 & 7.7\\
\hline
-4.95 & \multicolumn{1}{c}{4.95} & 26.8 & 27.2 & 1.2\\
\hline
-4.95 & \multicolumn{1}{c}{-4.95} & 26.6 & 27.4 & 3.0\\
\hline
0 & \multicolumn{1}{c}{3.0} & 24.4 & 26.1 & 7.2\\
\hline
0 & \multicolumn{1}{c}{-3.0} & 24.6 & 26.6 & 7.9\\
\hline
3.0 & \multicolumn{1}{c}{0} & 24.5 & 26.7 & 8.7\\
\hline
-3.0 & \multicolumn{1}{c}{0} & 24.9 & 26.1 & 4.8\\
\hline
2.83 & \multicolumn{1}{c}{2.83} & 24.5 & 26.8 & 9.4\\
\hline
2.83 & \multicolumn{1}{c}{-2.83} & 25.2 & 27.5 & 9.1\\
\hline
-2.83 & \multicolumn{1}{c}{2.83} & 25.0 & 26.2 & 5.1\\
\hline
-2.83 & \multicolumn{1}{c}{-2.83} & 25.1 & 26.2 & 4.2\\
\hline

\end{tabular}
\end{table}
The maps of the source images for different off-axis angle are shown in Figures \ref{subFig:PSFMAPMo}, \ref{subFig:PSFMAPTi} and \ref{subFig:PSFMAPFe} for 2.3 keV, 4.5 keV and 6.4 keV, respectively, while
the corresponding models of the PSFs, based on Equation \ref{eq:PSF}, are shown in Figures \ref{subFig:PSFModelMo}, \ref{subFig:PSFModelTi} and \ref{subFig:PSFModelFe}.
\begin{equation}
    \label{eq:PSF}
    PSF(r) = W e^{-\frac{r^{2}}{2\sigma^{2}}} + N \times (1+(\frac{r}{r_c})^2)^{-eta}+M\times(r+Offset)^{index}
\end{equation}

The Encircled Energy Function (EEF) is defined as:
\begin{equation}
    \label{eq:EEF}
    EEF(r) = \int^r_0 PSF(r)2\pi r dr
\end{equation}
and the HPD is defined as HPD = 2 $\times$ R where EEF(R) = 0.5 $\times$ EEF(r$_{max}$) with r$_{max}$ = 8 mm.

\subsubsection{Discussion}\label{sec:HPDDisc}
The tables \ref{tab:HPD2.3keV}--\ref{tab:HPD6.6keV} show that the predicted and measured telescope angular resolutions are in reasonable agreement. At each energy, taking the dataset as a whole, there is a small offset (measured minus predicted) in resolution plus a statistical fluctuation around that offset. For all energies, the statistical fluctuations are around 2-3\% (1-$\sigma$), consistent with the estimated measurement uncertainties (see Table captions). The offset in HPD values is -2.1\% for 2.3 keV, 1.4\% for 4.6 keV, and 5.7\% for 6.6 keV. The fact that there is a very slight underestimate at lower energies and a slight overestimate (amounting to about 1.5 arcsec) at higher energies remains a mystery. It should be noted that, despite these small differences, the on-axis measured telescope angular resolution in flight is still well within requirements at all energies. Since launch, the angular resolution of the flight system (3 telescopes combined) has been regularly measured using point sources and still (nearly 3 years after launch) meets its system level requirement ($\le$ 30 arcsec HPD).
Figures \ref{subFig:PSFMAPMo}, \ref{subFig:PSFMAPTi} and \ref{subFig:PSFMAPFe} illustrate that the PSF is well behaved even near the edge of the field-of-view, which is essential when mapping polarization from extended sources. 
 
\section{Conclusion}
The principal goal of the telescope on-ground calibration, carried out using flight-spare units, was to demonstrate that, to the accuracy required in the calibration plan, the performance of each flight telescope could be synthesized from the substantial amount of calibration data that were collected on the individual flight mirror modules and flight detector units, in the U.S and Italy, respectively.  Areas of concern were that detector calibrations were performed with nominally parallel beams of X rays while near the focus of an optic, the input beam is a converging, then diverging cone of X rays. This could potentially affect the modulation factor, and the spurious modulation inherent in all the detectors, rendering their calibrations inaccurate. It can also have an effect on the spatial resolution of the detector, due to the transparency of the fill gas and the finite detector depth, which must be modeled accurately to incorporate. Finally, X-ray reflection can affect the polarization measured from a source, although this effect is calculated to be very small for typical graze angles.

The data taken here show that the inclusion of the mirror module does not affect the polarization response of the detectors within the accuracy required for IXPE calibration and that the telescope angular resolution can be determined. Thus, the original stand-alone calibrations of the flight MMAs and the flight detector units can be used in full to derive the telescope calibrations. 

\bibliography{References}{}
\bibliographystyle{aasjournal}
\end{document}